\documentstyle[epsfig,floats,aps]{revtex}

\begin{document} 

\draft

\title{A method to determine the polarization of high energy gamma - rays}
\author{Gerardo O. Depaola\thanks{Electronic address: depaola@fis.uncor.edu}, Carlos N. Kozameh\thanks{Electronic address: kozameh@fis.uncor.edu} and Manuel H. Tiglio\thanks{Electronic address: tiglio@fis.uncor.edu} }
\address{\it Facultad de Matem\'{a}tica, Astronom\'{\i}a y F\'{\i}sica, Universidad Nacional de C\'{o}rdoba. Ciudad Universitaria. 5000 C\'{o}rdoba, Argentina}						
\maketitle

\begin{abstract}
A method to determine the polarization direction of a very high energy (50 MeV - 30 GeV), linearly polarized gamma ray is analyzed and discussed. We study the differential cross section for pair production and show that the only relevant piece of information comes from the reconstruction of the pair trajectory. In particular the polarization information is coded in the azimuthal distribution of the created pair. We show that the Azimuthal Ratio (AR), defined as the ratio of produced pairs parallel to the polarization vector to pairs perpendicular to this vector, yields a very practical method to determine incoming polarization. We investigate the behavior of this function for different angular resolutions of the measuring apparatus and different energies of the incoming pairs.
\end{abstract}


\section{Introduction}
The study of high energy gamma rays produced by astrophysical sources give us a unique window to study very unusual objects in our universe. The GRO \cite{fit} has provided plenty of new data to study gamma ray bursts, and to obtain detailed maps of our galaxy in several energy ranges \cite{sch}. At present there are several projects to improve the resolution of GRO \cite{atw,dep} and give us a more accurate description that will enable us to decide if the gamma ray sources are extended or compact objects. It is worth mentioning that neither GRO nor the new projects have been designed to detect the polarization of those rays. However, one knows that polarized gamma rays are very important in astrophysics since they constitute a distinctive byproduct of matter accretion by a black hole or neutron star. Thus, it could be desirable to measure incoming polarization together with the other variables and perform a more comprehensive study of those astrophysical processes.

\begin{figure}
\centerline{\psfig{figure=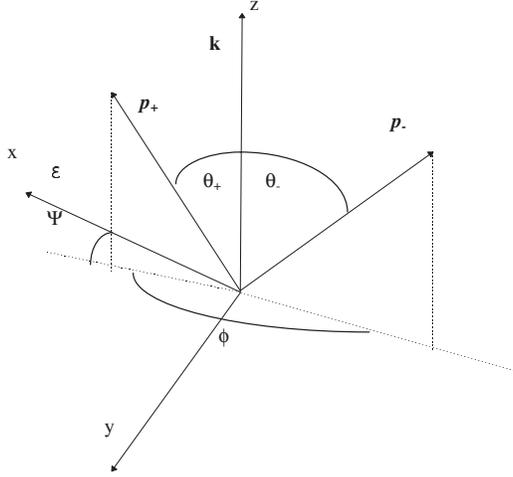,width=7cm}}
\caption{Angles occurring in the pair creation}
\label{fig1}
\end{figure}

We address here the problem of measuring the polarization of gamma rays in the energy range $50$ MeV - $30$ GeV. The interaction of gamma rays with matter in this range is dominated almost exclusively by pair production. Thus, the detection of those rays is based on measuring the electron - positron pair created by these photons (Fig.\ref{fig1}).

In this work we extensively study the cross section of pair production looking for what characteristics of this process can be useful to determine the polarization of the gamma - ray. Since most astrophysical processes of interest generate linearly polarized rays we well assume that the incoming ray is polarized along a definite direction (along the $x$ - direction in Fig.\ref{fig1}).

We first show that it is not possible to determine the incoming polarization by measuring the spin of the resulting pair. This leaves the study of the particle track as the only mean to detect the polarization direction. In particular, the polarization information lies in the azimuthal distribution of the particles moments so we calculate this probability distribution integrating the differential cross section in energy and polar angles.

The idea to detect polarization of a gamma ray by measuring the azimuthal distribution of the created pair was brought up in the $50$'s. It was then determined that the asymmetric ratio (AR) of the number of pairs produced in a plane parallel to the polarization vector, to the number perpendicular to that direction is a good parameter to describe this distribution \cite{ber,wic}. Although at that time only cross sections for coplanar events were used, Maximon and Olsen later showed that this ratio is also useful when the events make a small but finite angle $\Delta \epsilon$ around the coplanar direction \cite{max}.

However, since the differential cross section for this process is a very complicated function of the relevant angles, the above mentioned authors used approximations with different, and sometimes opposite, results. As an example, Berlin et al. \cite{ber} calculated a higher probability of pair production perpendicular to the polarization vector, whereas Wick \cite{wic} obtained the opposite result. Maximon and Olsen \cite{max}, using quantum mechanical calculations, found that the azimuthal relation asymmetry depended on the angular resolution with which one measures these probabilities. Thus, depending on this resolution, one might have a higher probability of pair production in either direction. However, since their calculations were only done for small deviations from coplanarity it was not clear what would be the behavior of AR as a function of the azimuthal angle.

It is worth mentioning that none of these authors computed the azimuthal probability distribution which is necessary to perform Monte Carlo simulations test to the prototypes.

In this work we calculate this spatial azimuthal distribution for high energy gamma rays (50 MeV - 30 GeV) directly from the differential cross section by integrating over the polar angles and the positron (electron) energy.

The integration is made done using a Monte Carlo method and the VEGAS routine of Numerical Recipes \cite{num}. To check the method, we calculate the total cross section and find that it yields the tabulated values.

In Section II we first compute the scattering cross section for pair production {\em without} summing over spins or polarization directions. This would be a standard problem for a course on QED but it is not discussed on any textbook on the subject. The main result obtained is that, in the high energy limit, the spinorial part factorizes out in the final expression thus showing that the polarization angle does not mix with the spin directions. We then sum over the spin directions obtaining the standard May formula \cite{jau}.

In Section III we compute the spatial azimuthal distribution by integrating the cross section over the energy and polar angles. We also parametrize this distribution with a `fitting function' of the two azimuthal angles.

Next, in Section IV, we compute the asymmetric ratio AR as a function of the indeterminacy in the azimuthal direction. This is done by integrating over the angle $\Delta \epsilon$ and an appropriate energy range of the produced pair. Calculations are repeated for different incoming energies of the gamma ray.

Finally, in Section IV we argue that the best method of detecting incoming polarization is to select the angular resolution that yields the maximum value of AR.

\section{Relativistic cross section for linearly polarized gamma ray}

The pair production by a high energy incident photon can be analyzed with a very good approximation by assuming the incident photon interacts with an external potential, which is assumed to be a spherically symmetric, static one (Born approximation). In the following, $u_{r}$ and $v_{s}$ ($r,s=1,2$) denote spinors for the electron and positron respectively, such that $\Pi ^{r} u_{r'} = \delta _{rr'} u_{r}$, $\Pi ^{r} v_{r'} = (1-\delta _{rr'}) v_{r}$, where $\Pi ^{1}$ ($\Pi ^{2}$) is the positive (negative) helicity projection operator; and $\Lambda ^{+}$ ($\Lambda ^{-}$) is the positive (negative) energy projection operator. That is,
$$
\Pi^{r}(p) = \frac{1}{2}\left[ 1+(-1)^r \sigma _{p}  \right]  \;\;\;\;\; , \;\;\;\;\; \Lambda^{\pm}(p) = \frac{1}{2m} \left( \pm \not \! p +m  \right) \; .
$$

The Feynman amplitude for a pair defined by these spinors is ${\cal M}={\cal M}_{a} + {\cal M}_{b}$, with
\begin{displaymath}
{\cal M}_{a}=-ie^{2}\bar{u}_{r}(\vec{p_{1}}){\not \! \epsilon} S(p'_{a})\not \! \! A(\vec{q})v_{s}(\vec{p_{2}}) \; \; \; \; , \; \; \; \; {\cal M}_{b}=ie^{2}\bar{u}_{r}(\vec{p_{1}})\not \! \! A(\vec{q})S(p'_{b}){\not \! \epsilon}v_{s}(\vec{p_{2}}) \; ,
\end{displaymath}
and
\begin{displaymath}
\not \! \! A(\vec{q})=\frac{eZ{\gamma^{0}}}{\left| \vec{q} \right| ^{2}} \; \; \; \; \; \; , \; \; \; \; \; \; S(p'_{i})=\frac{\not \! \! p'_{i}+m}{(p'_{i})^{2}-m^{2}} \; \; \; \; \; \; \; (i=a,b) \; .
\end{displaymath}
The absolute value of ${\cal M}$,
$$
\left| {\cal M} \right| ^{2} = \left| {\cal M}_{a} \right| ^{2} + \left| {\cal  M}_{b} \right| ^{2} + 2 \Re{( \bar{\cal M}_{a}{\cal M}_{b} )} \; ,
$$
can be computed with the aid of the above mentioned projectors. Defining
$$
T_{a} \equiv \not \! \epsilon (\not \! p_{1}-\not \! k + m) \not \! \gamma ^{0} \;\;\;\; , \;\;\;\;\; T_{b}  \equiv  \not \! \gamma^{0} (\not \! p_{2}-\not \! k+m) \not \! \epsilon \;\;\;\;\; , \;\;\;\;\; c_{i} \equiv   \frac{e^{3}Z} {\left| \vec{q} \right| ^{2} \left[ (p'_{i})^{2}-m^{2} \right] } \; \; \; \; \; \; \; \; \; \; \; \; \; \; \; \;  (i=a,b) \; ,
$$
using the identity $u_{r}(p) \bar{u}_{r}(p) = \Pi ^{r} \Lambda ^{+}(p)$ and similarly for $v$, we find that
\begin{eqnarray}
\left| {\cal M}_{a}\right| ^{2} & = &  - c_{a}^{2} \: tr [ \: \Lambda ^{+} (p_1) \: T_{a} \: \Pi ^{s} \: \Lambda ^{-} (p_2) \: T^{\dagger}_{a} \: \Pi ^{r} ]  \; , \label{tr1}\\
\left| {\cal M}_{b}\right| ^{2} & = & - c_{b}^{2} \: tr [ \: \Lambda ^{+} (p_1) \: T_{b} \: \Pi ^{s} \: \Lambda ^{-} (p_2)\: T^{\dagger}_{b} \: \Pi ^{r} ] \; ,  \label{tr2}\\
\Re{ ( \bar{\cal M}_{a}{\cal M}_{b} ) } & = & - c_{a}c_{b} \: \Re{ \left\{ tr [ \: \Lambda ^{+} (p_1)  \: T_{b} \: \Pi ^{s} \: \Lambda ^{-} (p_2)  \: T^{\dagger}_{a} \: \Pi ^{r} ] \right\} } \label{tr3}\;  .
\end{eqnarray}

Since we are interested in very energetic gamma rays, we can take the ultrarelativistic limit for the projection operators, namely,
$$
\Lambda ^{+} (p_1) \approx  \frac{\not \!p_1}{2m} \;\;\; , \;\;\; \Lambda ^{-} (p_2)  \approx  \frac{\not \!p_2}{2m} \;\;\; , \;\;\;
\Pi ^{^q}  \approx  \frac{1}{2} [ 1+ (-1)^q \gamma ^5 ] \; .
$$
Inserting the above operators in Eqs.(\ref{tr1},\ref{tr2},\ref{tr3}) and recalling that no linear terms in $\gamma ^5$ must appear in the final expressions since we are dealing with real quantities, we obtain
\begin{eqnarray*}
|{\cal M}_{a}|^{2} & = & \frac{c_{a}^{2}}{16m^{2}}[1+(-1)^{r+s}] \; tr [ \not \! p_{1}{\not \!\epsilon} (\not \! p_{1}- \not \! k) {\not \! \gamma^{0}}\not \! p_{2}{\not \! \gamma^{0}}(\not \! p_{1}-\not \! k){\not \! \bar{\epsilon}} ] \; , \\
|{\cal M}_{b}|^{2} & = &\frac{c_{b}^{2}}{16m^{2}}[1+(-1)^{r+s}] \; tr [ \not \! p_{2}{\not \! \epsilon}(\not \! p_{2}-\not \! k){\not \! \gamma^{0}} \not \! p_{1}{\not \! \gamma^{0}}(\not \! p_{2}-\not \! k){\not \! \bar{\epsilon}} ] \; ,  \\
\Re{( \bar{\cal M}_{a} {\cal M}_{b} )}& = & - \frac{{c_{a}}c_{b}}{16m^{2}}[1+(-1)^{r+s}] \; tr [ \not \! p_{1}{\not \! \gamma^{0}}(\not \! p_{2}-\not \!k) {\not \! \epsilon}\not \! p_{2}{\not \! \gamma^{0}}(\not \! p_{1}-\not \! k){\not \! \bar{\epsilon}} ] \; . \\
\end{eqnarray*}
Note that the spinor contribution factorizes out in the three equations, and that only two final states  (which are equally probable) are allowed, independently of photon's polarization. Thus, it is not possible to know the latter from the spins of the pair.
We now proceed by adding over the two possible final states. Assuming that the photon is lineally polarized we obtain
\begin{eqnarray}
{| \cal M |}^{2} & = & - \frac{e^6 Z^2} {2m^{2}|\vec{q}|^{4}}  \left[   \frac{(\epsilon p_{1})^{2}} {(p_{1}k)^{2}}  \left(4E_2^2 - |\vec{q}|^2 \right)  +  \frac{(\epsilon p_{2})^{2}} {(p_{2}k)^{2}} \left(4E_1^2- |\vec{q}|^2 \right) \right. \nonumber \\
& & \left. + \frac{2(p_1\epsilon )(p_2\epsilon )} { (p_1k) (p_2k) } \left( 4E_1E_2 - |\vec{q}|^2 \right) + 2 + \frac{(p_2\epsilon)}{(p_1k)} + \frac{(p_1\epsilon)}{(p_2k)} - \frac{2 \omega ^2 |\vec{q}|^2}{(p_1k)(p_2k)}  \right]  \; .    \label{ultrarel}
\end {eqnarray}

The cross section for pair production by lineally polarized gamma rays in the high energy or ultrarelativistic limit is thus obtained by inserting the Feynman amplitude (\ref{ultrarel}) in
$$
d \sigma = \left( 2 \pi \right)^2 \frac{\alpha Z^2}{4 \pi} r_0 \int \frac{m^4}{E_+ E_- \omega |\vec{q}|^4 \left( 2m \right)^2} |{\cal M}|^2 d^3 p_1 d^3 p_2 \delta\left( E_+ + E_- - \omega \right) \; .
$$
Computing the scalar products and using natural units with $h/2\pi =c=1$ this gives
\begin{eqnarray}
d\sigma & = & \frac{-2 \alpha Z^2r_0m^2}{(2 \pi)^2 \omega ^3} dE d\Omega _+ d\Omega_ - \frac{E(\omega - E)}{|\vec{q}|^4} 
\left\{ 4 \left[ E \frac{\sin{\theta _-} \cos{\Psi}}{1-\cos{\theta _-}} + (\omega - E) \frac{\sin{\theta _+} \cos{(\Psi + \phi)}} {1- \cos{\theta _+}}  \right] ^2 \right.   \nonumber \\
& &  - |\vec{q}|^2 \left[ \frac{\sin{\theta _-}\cos{\Psi}}{1-\cos{\theta _-}} - \frac{\sin{\theta _+ \cos{(\Psi + \phi)}}}{1- \cos{\theta _+}} \right] ^2  \nonumber \\
& & \left. - \omega ^2 \frac{\sin{\theta _-} \sin{\theta _+}}{(1-\cos{\theta _-})(1- \cos{\theta _+})} \left[ \frac{E\sin{\theta _+}}{(\omega _ E)\sin{\theta _-}} + \frac{(\omega - E)\sin{\theta_-}}{E \sin{\theta _+}} + 2\cos{\phi} \right] \right\} \; , \label{eficaz}
\end{eqnarray}
with
$$
|\vec{q}|^2 = -2 \left[ E(\omega - E)(1 - \sin{\theta _+} \sin{\theta _-} \cos{\phi} - \cos{\theta _+} \cos{\theta _-}) + \omega E(\cos{\theta _+}-1) + \omega (\omega - E)(cos{\theta _-}-1) + m^2 \right]  \; .
$$
$E$ is the positron energy and we have assumed that the polarization direction is along the $x$ axis (see Fig.\ref{fig1}).

\section{Spatial azimuthal distribution}

Integrating Eq.({\ref{eficaz}) over energy and polar angles yields the spatial azimuthal distribution. Since an analytic integration is almost impossible to perform without some sort of approximation, we numerically computed the integral using a Monte Carlo procedure.

Fig.\ \ref{fig2} shows an example of this distribution for $100$ MeV gamma - ray. In this figure the range of the $\phi$ axis is restricted between $3.0$ and $\pi$ since it gives the most interesting part of the distribution. For angles smaller than $3.0$ this distribution monotonically decreases to zero.

\begin{figure}[h]
\centerline{\psfig{figure=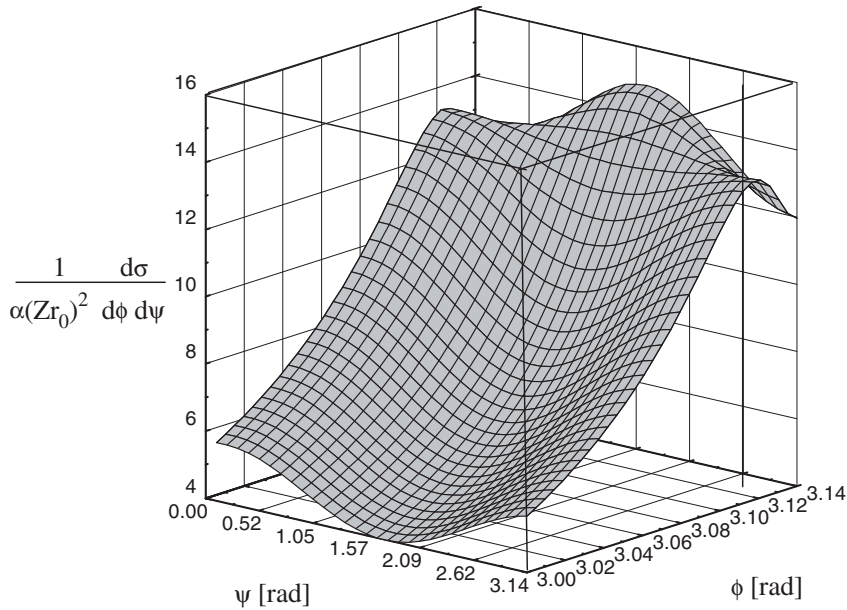,width=12cm,angle=-90,clip=}}
\label{fig2}
\caption{Spatial azimuthal distribution of a pair created by $100$ MeV photon}
\end{figure}

We also note that for exactly coplanar events (that is, for $\phi = \pi$) it is more probable to find a pair in a plane perpendicular to the polarization vector that in a plane parallel to this vector, but this tendency changes rapidly as $\phi$ decreases. Repeating this calculation for different values of energies of the incoming ray yields similar results.

Since one of our goals is to use the azimuthal distribution in numerical simulations of the performance of a gamma ray detector, it is very convenient to parametrize the distribution surface in terms of smooth functions of $(\phi , \psi )$ . This parametrization is done as follows. We first select coplanar events, $\phi = \pi$. Instead of using a fitting function $A^* (1- \lambda ^* \cos ^2 \psi )$, used in previous works \cite{dep2}, we choose a form $f(\pi ,\psi ) = f_{\pi/2} \sin ^2 \psi + f_0 \cos ^2 \psi$ (note that the asymmetric ratio can be computed by the quotient $f/f_{\pi /2}$). As we move off this coplanar situation, $f_0$ and $f_{\pi /2}$ become functions of $\phi$, that is:

\begin{equation}
f(\phi ,\psi ) = f_{\pi /2}(\phi ) \sin ^2 \psi + f_0 (\phi) \cos ^2 \psi \; . \label{ajuste}
\end{equation}
Thus, to parametrize the entire surface we simply evaluate the azimuthal distribution for two values of $\psi$, namely, $0$ and $\pi /2$ .

Since both $f_0 (\phi)$ and $f_{\pi /2} (\phi)$ are functions that rapidly vary when $\phi$ approaches $\pi$, it was necessary to adjust the functions in two ranges of $\phi$: (I) $0 \leq \phi \leq 3.05$ rad. (II) $3.06$ rad $\leq \phi \leq \pi$ , whereas in the small range $3.05 \leq \phi \leq 3.06$ we extrapolate the two fitting functions until the intersection point is reached (this minimizes the errors in the above mentioned range).

In region I we used Lorentzian functions of the form
\begin{equation}
f(\phi ) = y_0 + \frac{2A \omega}{\pi [\omega ^2 + 4 (\phi - x_c)^2]} \; ,
\end{equation}
whereas for region II the best fitting function was found to adopt the form:
\begin{equation}
f(\phi ) = a + d \tan{(b \phi + c)} \; .
\end{equation}

Table \ref{tabla1} and \ref{tabla2} show the results obtained.

\begin{table}
\caption{Parameters for the Lorentzian fitting $f_0(\phi )$ and $f_{\pi /2}(\phi )$ .}
\begin{center}
\begin{tabular}{ c c  c c c c c c c} 
Lorentzian & & $f_0 (\phi)$ & & & & $f_{\pi /2}(\phi)$ &  & \\ \hline
Energy & $y_0$ & A & $\omega$ & $x_c$ & $y_0$ & A & $\omega$ & $x_c$ \\ \hline
$100$ MeV & $6.0$ & $1.03$ & $0.089$ & $3.118$ & $3.4$ & $1.84$ & $0.101$ & $3.146$ \\
$200$ MeV & $8.16$ & $1.76$ & $0.0598$ & $3.130$ & $4.7$ & $2.7$ & $0.064$ & $3.148$ \\
$500$ MeV & $10.8$ & $3.06$ & $0.036$ & $3.138$ & $6.3$ & $4.8$ & $0.034$ & $3.150$ \\
$700$ MeV & $12$ & $3.6$ & $0.029$ & $3.139$ & $6.9$ & $6.0$ & $0.026$ & $3.15$ \\
$1$ GeV & $13$ & $4.3$ & $0.0248$ & $3.141$ & $7.8$ & $7.8$ & $0.018$ & $3.15$ \\
\end{tabular} 
\end{center}
\label{tabla1}
\end{table}

\begin{table}
\caption{Fit for the parameter of $f_0 (\phi )$ function.}
\begin{center}
\begin{tabular}{c c c  c  c}
Parameter & Function & a & b & c \\ \hline
$y_0$ & $a\ln{E} - b$ & $2.98 \pm 0.06$ & $7.7 \pm 0.4$ & - \\ 
$A$ & $a\ln{E} - b$ & $1.41 \pm 0.08$ & $5.6 \pm 0.5$ & - \\ 
$\omega$ & $a + b/E + c/E^3$ & $0.015 \pm 0.001$ & $9.5 \pm 0.6$ & $(-2.2 \pm 0.1) 10^4$ \\
$x_c$ & $a + b/E + c/E^3$ & $3.143 \pm 0.001$ & $-2.7 \pm 0.2$ & $(2 \pm 1)10^3$ \\ 
\end{tabular}
\end{center}
\label{tabla2}
\end{table}

\begin{table}
\caption{Fit for the parameter of $f_{\pi /2} (\phi)$ function.}
\begin{center}
\begin{tabular}{c c c c c} 
Parameter & Function & a & b & c \\ \hline
$y_0$ & $a\ln{E} - b$ & $1.85 \pm 0.07$ & $5.1 \pm 0.4$ & - \\ 
$A$ & $a\ln{E} - b$ & $1.3 \pm 0.1$ & $(6.6 \pm 0.2)10^{-3}$ & - \\ 
$\omega$ & $a + b/E + c/E^3$ & $0.008 \pm 0.002$ & $12.1 \pm 0.9$ & $(-2.8 \pm 0.8) 10^4$ \\ 
$x_c$ & $3.149$ & - & - & - \\ 
\end{tabular}
\end{center}
\label{tabla3}
\end{table}

It is interesting to observe that for region I the difference between the calculated distribution (Fig.\ \ref{fig2}) and the parametric form given by (\ref{ajuste}) with the parameters of Table \ref{tabla1} is less than $5 \%$. Note also that the coefficients themselves vary smoothly over the energy range. It is thus possible to give a parametric form of these coefficients as functions of the incoming energy. In Tables \ref{tabla2} and \ref{tabla3} we give the explicit forms of $\gamma _0 (E)$ , $\omega (E)$ , $A(E)$ and $x_c (E)$ associated with $f_0 (\phi)$ and $f_{\pi /2} (\phi)$ with appropriate fitting functions.

The parametric description of Region II is given below in Table \ref{tabla4}. It follows from this table that the parameters associated with the tangent function are approximately the same for $f_0 (\phi)$ and do not differ by much for $f_{\pi /2} (\phi)$ . Thus, if we consider that this part of the azimuthal function is very small when compared with the range close to $\pi$ , we find it reasonable to take an average of these constants for the entire energy range.

\begin{table}
\caption{Parameters for the tangent fitting function $f_0 \left(\phi \right)$ and $f_{\pi /2}\left(\phi \right)$}
\begin{center}
\begin{tabular}{c c c c c c c c c} 
Lorentzian & & & $f_0 \left(\phi \right)$ & & & $f_{\pi /2} \left(\phi \right)$ & &  \\ \hline
Energy & a & b & c & d & a & b & c & d \\ \hline
$100$ MeV & $-0.21$ & $-0.287$ & $2.48$ & $-0.29$ & $0.021$ & $-0.271$ & $2.425$ & $-0.16$ \\
$200$ MeV & $-0.19$ & $-0.289$ & $2.48$ & $-0.28$ & $0.056$ & $-0.288$ & $2.47$ & $-0.16$ \\ 
$500$ MeV & $-0.19$ & $-0.289$ & $2.48$ & $-0.28$ & $0.054$ & $-0.286$ & $2.47$ & $-0.17$ \\ 
$700$ MeV & $-0.19$ & $-0.289$ & $2.48$ & $-0.28$  & $0.04$ & $-0.281$ & $2.45$ & $-0.17$ \\
$1$ GeV & $-0.19$ & $-0.281$ & $2.45$ & $-0.27$ & $0.04$ & $-0.278$ & $2.439$ & $-0.16$ \\
\end{tabular} 
\label{tabla4}
\end{center}
\end{table}

\section{Determining gamma ray polarization with the asymmetric ratio}
From the azimuthal probability distribution we define the asymmetric ratio (AR) as follows. We first select the azimuthal distribution at $\phi =\pi$ (coplanar events), two azimuthal angles $\psi = 0,\pi /2$ (parallel and perpendicular to the polarization direction, respectively) and numerically integrate the distribution around those values of $(\phi ,\psi )$ in the range $[-\Delta \epsilon , \Delta \epsilon]$. The AR is defined as the number of pairs that are contained in a plane parallel to the polarization vector to the number in a plane perpendicular to that vector. Note that the integration is done on the two angles $(\phi , \psi )$ since we are assuming that $\Delta \epsilon$ comes from the angular resolution of the detector.

In Fig.\ \ref{fig3} we present the AR as a function of angular resolution for $100,500$ and $1000$ MeV gamma ray energy. From this figure one can see that the best values for AR are obtained forn angular resolutions between $0.2$ and $0.7$ rad, where the curve reaches a maximum value above $1.25$. Fig.\ \ref{fig3} shows that there are two possible ways to determine the incoming polarization. If one has outstanding angular resolution of the measuring apparatus and considers only coplanar pairs, one obtains an AR of approximately $0.8$, namely, the coplanar pairs will emerge perpendicular to the polarization direction. On the other hand, if the detector has poor angular resolution or if one considers emerging pairs in a wide band of angular directions around coplanarity, one finds that more pairs emerge parallel to the polarization direction with an AR better than $1.25$. Not only this value of the AR maximum has a greater deviation from one than $0.8$, but also there is a practical reason to select a coarse band of angular directions, to measure the $0.8$ value one needs a very good resolution while for $1.25$ one does not.

\begin{figure}[h]
\centerline{\psfig{figure=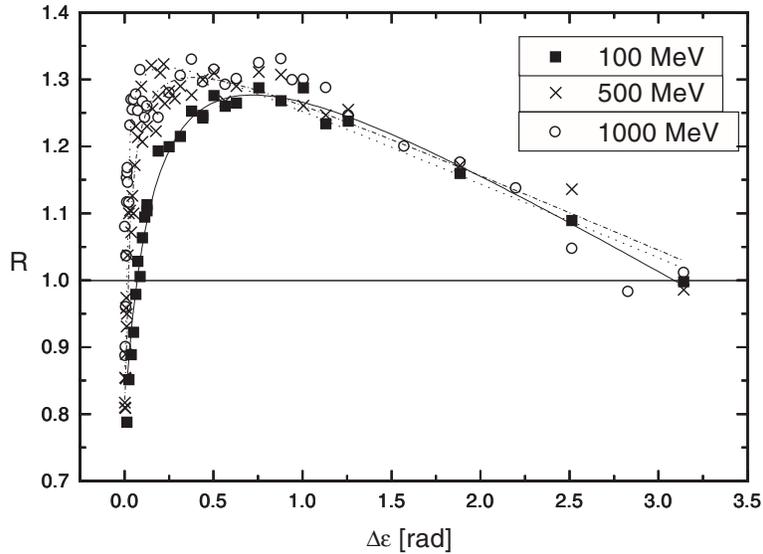,width=10cm,angle=90,clip=}}
\caption{Asymmetric ratio for several gamma - ray energies versus azimuthal resolution}
\label{fig3}
\end{figure}

\begin{figure}[h]
\centerline{\psfig{figure=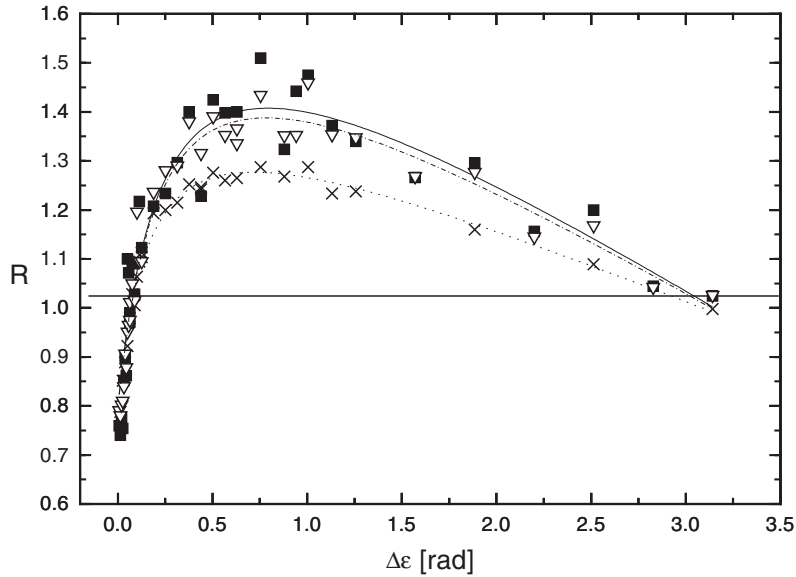,width=10cm,angle=90,clip=}}

\caption{Comparison of three asymmetric ratio for $100$ MeV incoming energy using different selection criteria. (Cross) without selecting energy; (Box) selecting particles with energy between $45$ and $55$ MeV; (Diamond) selecting between $40$ and $60$ MeV.}
\label{fig4}
\end{figure}

The reader should not conclude from the above paragraph that one should build a poor angular resolution detector (with great savings for the project) to measure polarized gamma rays. The analysis was done under the assumption that there was only one incoming polarization direction. If one has astronomical sources that produce two or more polarization directions one obviously needs higher resolution to resolve these directions.

The values of AR obtained in Fig.\ \ref{fig3} can be improved if one selects particles with some specific values of energy. In particular, if one selects particles with energies close to equipartition of the incoming photon energy one can improve the value of the asymmetric ratio.

In Fig.\ \ref{fig4} we compare three AR for incoming $100$ MeV gamma rays. The lowest curve is obtained by selecting all possible energies, the second curve is obtained by selecting particles with energies $\omega /2 \pm 0.05 \omega$, whereas the last one has energies in the range $\omega /e \pm 0.1 \omega$.

Note that when one carefully selects the energy range of the electron - positron pairs, one improves the values of AR achieving a maximum of approximately $1.4$ at $0.65$ rad.

On the other hand, this improvement of $10 \%$ is obtained at the expense of discarding more than $80 \%$ of the produced pairs. Thus, it would appear that to determine the AR it is more reasonable to simply integrate over the whole energy scale.

 \section{Summary and conclusions}
In this work we have explicitly computed the differential cross section for pair production and the azimuthal distribution of the produced pairs. We have also given a parametric form of this distribution in the different regions of interest. This parametric form is specially useful when implementing a Monte Carlo routine. We have then shown that the Asymmetric Ratio yields, in principle, a method to detect the polarization direction of the incoming photon. It is interesting to note that the azimuthal distribution surface changes rather abruptly for different values of $\phi$ near coplanarity. The AR function reflects this behavior. It starts at a value of $0.8$ for coplanar events and reaches a maximum value above $1.25$ when the angular distribution is about $0.5$ rads. The graph of AR explains and resolves the discrepancy between different authors who used this function to detect polarization direction.

As we mentioned before the amplitude of the AR can be improved if one selects the energy range of the produced pairs. However, this improvement of approximately $10 \%$ is obtained at the expense of discarding more than $80 \%$ of the pairs. It then appears that unless the intensity of the incoming gamma rays is very high, it is more convenient to use the entire energy range of the outgoing pairs.

The main conclusion one draws from the AR is that it is not necessary to have a detector with very good angular resolution to determine whether or not the incoming rays are linearly polarized. On the other hand, one needs very good resolution to estimate the direction of the polarization vector. Thus, any `reasonable' resolution that enables to determine the direction of the incoming photon \cite{dep} should be adequate to then decide whether or not he photon is linearly polarized. The better angular resolution of the detector, the more accurate the determination of the incident direction and polarization vector of the gamma ray.

Finally, a question we would like to address is whether or not it is possible to construct a detector capable of measuring the incident polarization of gamma rays produced by astrophysical sources. Preliminary results obtained with a monte carlo code simulation (where it was assumed that all events are coplanar) show that it is possible to determine the gamma ray polarization with an optimal geometrical setup\cite{dep2} . Future work will include the parametric form of the azimuthal distribution here discussed for a more realistic monte carlo simulation. We will also analyze alternative geometries of the measuring apparatus to minimize the errors introduced by multiple scatterings of the produced pairs. 

\section*{Acknowledgements}
This work was partially supported by grants from CONICET, CONICOR, and the National University of C\'ordoba. GOD and CNK acknowledge financial support from AIT, and MHT from CONICOR.

\end{document}